\documentclass[aps,prl,twocolumn,nofootinbib,longbibliography,superscriptaddress]{revtex4-1}

\usepackage{enumerate}
\usepackage{fancyhdr}
\usepackage{amsmath,amsfonts,amssymb, dsfont}
\usepackage{empheq}
\usepackage[colorlinks,citecolor=blue,linkcolor=blue,urlcolor=blue]{hyperref}

\usepackage{color}
\definecolor{red}{rgb}{1,0,0}
\definecolor{gre}{rgb}{0,0.6,0}
\definecolor{blu}{rgb}{0,0,1}

\usepackage{graphicx}

\begin{document}

\title{Four-dimensional Quantum Gravity with a Cosmological Constant from Three-dimensional Holomorphic Blocks}

\author{Hal M. Haggard}
\email{hhaggard[at]bard[dot]edu}
\affiliation{Physics Program, Bard College, 30 Campus Rd, Annandale-on-Hudson, NY 12504, USA}
\author{Muxin Han}
\email{hanm[at]fau[dot]edu}
\affiliation{ Department of Physics, Florida Atlantic University, 777 Glades Rd, Boca Raton, FL 33431, USA}%
\affiliation{Institut f\"ur Quantengravitation, Universit\"at Erlangen, Staudtstrasse 7, D-91058 Erlangen, EU}
\author{Wojciech Kami\'nski}
\email{wojciech.kaminski[at]fuw.edu[dot]pl}
\affiliation{Instytut Fizyki Teoretycznej, Uniwersytet Warszawski, ul. Pasteura 5, 02-093 Warszawa, Poland}
\author{Aldo Riello}
\email{ariello[at]perimeterinstitute[dot]ca}
\affiliation{Perimeter Institute for Theoretical Physics, 31 Caroline St N, Waterloo, Ontario N2L2Y5 Canada}

\begin{abstract}
\noindent  Prominent approaches to quantum gravity struggle when it comes to incorporating a positive cosmological constant in their models.  Using quantization of a complex $\mathrm{SL}(2, \mathbb{C})$ Chern-Simons theory we include a cosmological constant, of either sign, into a model of quantum gravity. 
\end{abstract}

\maketitle

The universe is accelerating in its  expansion. The simplest consistent explanation for the empirical data is a positive cosmological constant $\Lambda = 10^{-52}$ m$^{-2}$. A common expectation is that a quantum theory of gravity will shed light on the surprising value of this constant \cite{Weinberg:1989,Carroll:2001,Martin:2012} and on the interplay between quantum fields and the bare value of $\Lambda$ \cite{Carlip:1997,
Verlinde:2000}. A solid evaluation of this proposal requires the inclusion of $\Lambda$ into quantum gravity.  

Prominent approaches to quantum gravity, such as  the AdS/CFT framework of string theory, covariant loop quantum gravity, and group field theories, struggle to incorporate a positive cosmological constant, if they are able to incorporate one at all. A few frameworks, like causal dynamical triangulations and asymptotic safety, do not have this difficulty. 

In this letter we present a novel model of four-dimensional Lorentzian quantum gravity based on a discretized path integral over gravitational holonomy variables
where the cosmological constant emerges geometrically as a consequence of our quantization procedure via complex Chern-Simons theory. In particular, both signs of $\Lambda$ are treated on an equal footing in the model. 

Our discretization of the path integral decomposes spacetime into a simplicial complex, with each simplex of constant curvature $\Lambda$. We choose to work with parallel transports along closed paths (holonomies) as they are the most natural gravitational observables. They also fit nicely with the use of constant curvature simplices, whose geometry, as shown below, can be completely encoded in a finite number of holonomies. From the Chern-Simons perspective, these holonomies arise as the non-contractible cycles of a 3-manifold obtained by removing a graph from the 3-skeleton of the simplicial complex. 

Before delving into the details of the model, a few comments on  discreteness, symmetry, and the continuum limit. Discreteness is first introduced into the model as a regularization tool, as in a lattice gauge theory. However, the lattice will be dynamical and fluctuating in a quantum mechanical fashion. We show below that to avoid ambiguities in the model the spectrum of the fluctuations must have a Planck-scale discreteness and be gapped. Hence, the presence and absence of a lattice site is also subject to quantum fluctuations. This is the quantum `creation' and `annhilation' of the grains of spacetime.  

Because of the quantum dynamical nature of these discrete grains, the problem of recovering general relativity in the continuum limit is intertwined with a detailed understanding of their collective semiclassical behavior.

We do not treat the continuum limit here, however we provide evidence for a connection with classical gravity. This is achieved by studying the small $\hbar$ asymptotics of the quantum amplitude of a single building block. In this limit, we recover the exponential of the Einstein-Hilbert action (with the appropriate Gibbons-Hawking-York boundary term) for a simplex of constant curvature $\Lambda$. This is the curved Regge action of simplicial gravity.   The bulk action $\frac{1}{16 \pi G} \int \sqrt{-g} (R-2 \Lambda)$ corresponds, on the simplicial side, to a term $\frac{1}{8 \pi G}\Lambda V$, where $V$ is the 4-volume of a Lorentzian and homogeneously curved 4-simplex. Intriguingly, both the 4-volume and the boundary term are a result of the stationary phase evaluation of the Chern-Simons action subjected to appropriate boundary conditions.

The use of curved, instead of flat, simplices as building blocks serves a twofold dynamical purpose. Firstly, the interior of each building block is a solution of the dynamical equations of gravity with a cosmological constant and, secondly, by gluing a collection of these blocks one can build global solutions with the correct continuum symmetries \cite{Bahr:2009}, that is, those of (A)dS. In this sense, these building blocks are maximally adapted to the symmetries of the problem and, as such, are promising for studying perturbations around a background that includes a cosmological curvature---a potentially important technique for addressing the smallness of $\Lambda$.  

A further consequence of using these building blocks is that their curvature $\Lambda$ enters as an infrared cutoff scale. Thus, in this model the cosmological constant  plays a dual role to the Planck area, the latter providing a natural ultraviolet cutoff.  Together these scales make the quantum theory of a fixed discretization finite.

The model presented in this letter has a rich set of relations with other physical and mathematical theories. It can be seen as a generalization of the Turaev-Viro \cite{Turaev:1992}, and hence Chern-Simons-Witten \cite{Witten:1988}, quantization of three-dimensional gravity. The physical picture draws heavily from loop gravity, especially in its covariant, spinfoam version.  On the other hand, the mathematical tools come largely from the connections that Chern-Simons theory has with knot polynomials and the volume conjecture. The idea of a relation between Chern-Simons theory and four-dimensional loop gravity with a cosmological constant dates back to the early days of these theories \cite{Kodama:1988, Kodama:1990, Smolin:1995,Markopoulou:1998,Smolin:2002}. String theorists have worked extensively on complex Chern-Simons theory and we see here potential for exchange of ideas between loops and strings. 

In particular, the holomorphic blocks that play a key role in defining the present model have an M-theory interpretation, and closely relate to supersymmetric gauge theory, as well as surface operators with junctions \cite{Beem:2014, Dimofte:2014,Han:2015,Gadde:2013,Chun:2015}. 
While not holographic in any standard sense our model does exhibit a rich interplay between three- and four-dimensional topological and gravitational theories (for recent work on this interplay in lower dimensions, see e.g. \cite{Verlinde:2014}).  
We hope that recent advances in these active fields will accelerate progress in quantum gravity, for example in treating the smallness and continuum limit problems. 

\emph{Geometry from holonomies}---Parallel transport along a Faraday-Wilson loop of the gravitational field results in a holonomy that encodes aspects of the curvature of the region bounded by the loop. Consider an homogeneously curved, geodetic 4-simplex in an (A)dS space, and the set of holonomies along loops encircling the simplex's faces. Choose these loops to share a common base point and be contained in the 1-skeleton of the simplex. Not all of these loops are topologically independent and hence the holonomies are subject to a set of algebraic constraints. Before displaying these constraints explicitly we move to a dual description of the simplex.

\begin{figure}[t] 
   \centering
   \includegraphics[height=1.3in]{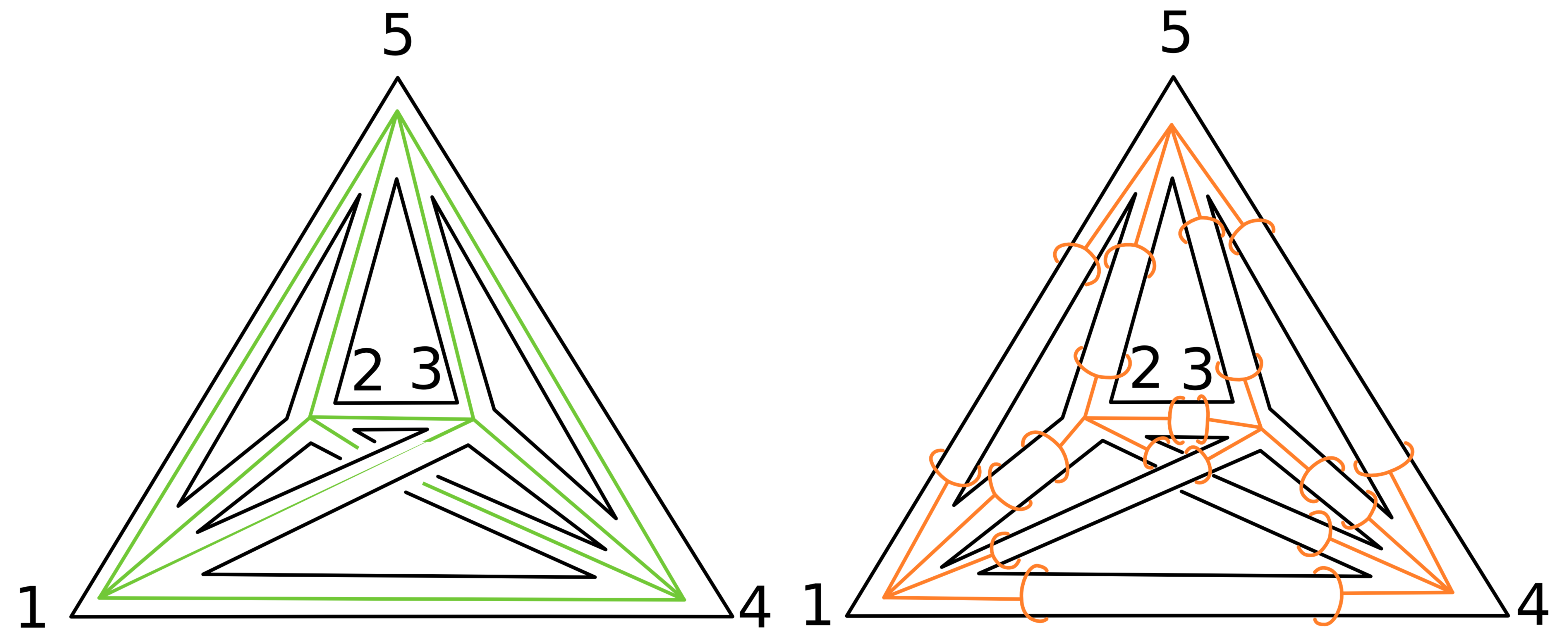} 
   \caption{A planar projection of the tubular neighborhood of the $\Gamma_5$ graph with the set of longitudinal and transverse paths used to calculate the holonomies $G_{ab}$ and $H_a(b)$, respectively. For simplices, the Poincar\'e dual is also a combinatorial simplex. }
   \label{graphs}
\end{figure}

The Poincar\'e dual of a 4-simplex is a graph, we call it $\Gamma_5$, that associates a vertex to each tetrahedron of the simplex and a graph edge to each pair of neighboring tetrahedra. These edges pierce the shared faces of the paired tetrahedra and hence are dual to these faces. Pictorially one can think of this as concentrating each of the faces on distributional defects (i.e. the edges of the dual graph) living in a three-sphere $S^3$ topologically equivalent to the boundary of the 4-simplex. With this image in mind it is clear that there is an isomomorphism between the fundamental group of the simplicial 1-skeleton and that of the graph-complement manifold $M_3 = S^3 \setminus N(\Gamma_5)$, where $N(\Gamma_5)$ is the open tubular neighborhood of the graph. 

Therefore the loops used to calculate parallel transports are in 1-to-1 correspondence with a set of equivalence classes of loops in the graph complement $M_3$. Associating the parallel transport holonomies to the corresponding equivalence class of loops we obtain a representation of a flat connection in $M_3$. The boundary of $M_3$, or equivalently of the closure of $N(\Gamma_5)$, is a genus six 2-surface, call it $\Sigma_6$. Any loop of $M_3$ can be deformed to live on this boundary. An (overcomplete) basis of non-contractible loops on $\Sigma_6$ is given by a set of $4 \times 5$ loops transverse to the tubes and 10 lines running longitudinally along the tubes, see Fig. \ref{graphs}. Labeling the vertices of $\Gamma_5$ with $a=1, \dots, 5$, the longitudinal holonomy from  $a$ to  $b$ is $G_{ba} = G_{ab}^{-1}$, and the transverse holonomy based at $a$ and encircling the tube $ab$ is $H_{b}(a)$. 

In this basis the topological constraints mentioned above break up into three classes. They are:
\begin{align}
\text{closures} & \quad \overleftarrow{\text{\Large $\Pi$}}_{ b } H_{b}(a) = \mathds{1}  ; \\
\hspace{-.2in} \text{parallel transports} & \quad G_{ba} H_b(a) G_{ab} = H_a(b)^{-1} ;\\
\text{bulk contractions} & \quad 
\begin{cases}
G_{ac} G_{cb} G_{ba} = \mathds{1} ,  \\
G_{34} G_{42} G_{23} = H_1(3) . 
\end{cases}
\label{eq_topoconstr}
\end{align}
The closures are a consequence of contractibility around each vertex. The parallel transport conditions are due to contractibility around each tube.  The last class of equations come from contractibility in the bulk of $M_3$ and are expressed in terms of a set of six independent cycles. The first five are $(abc) \in \{ 123, 125, 134, 235, 345 \},$ see Fig. \ref{graphs}, while the last is explicit. Eq. \eqref{eq_topoconstr} is a consequence of the single essential crossing in the planar projection.

Returning to these holonomies as gravitational observables on the 4-simplex we must account for the fact that each tetrahedron of the simplex defines a three-dimensional frame. This is why the four holonomies $\{H_b(a)\}_b$ associated to tetrahedron $a$ should stabilize the direction transverse to this frame (when appropriately parallel transported \cite{Haggardetal:2015, WHAM:2014,HHKR:2015}).  We take the transverse direction to be timelike and therefore these holonomies are part of an $\mathrm{SO(3)}$ subgroup of the Lorentz group. When viewed as a condition on the flat connection of $M_3$ we call this requirement the `geometricity constraint'.

We see that corresponding to each 4-simplex geometry in (A)dS there is a flat connection on $M_3$ satisfying the geometricity constraints. In a generalization of a classic theorem due to Minkowski \cite{Minkowski:1897,Aleksandrov:2005}, we have shown that the converse also holds: \\
{\bf Theorem 1.} \emph{There is a bijection between a dense set of Lorentz flat connections on $M_3$ satisfying the geometricity constraints and non-degenerate convex constant curvature 4-simplex geometries, up to parity. \cite{WHAM:2014,HHKR:2015}}\\
Because this is a bijection, a flat connection can be used to reconstruct either a positively or a negatively curved geometry, but only one of the two. For a precise statement of the non-degeneracy requirements, see \cite{Haggardetal:2015, WHAM:2014,HHKR:2015}. 

\emph{Sketch of the proof:} The $\mathrm{SO}(3)$ little group associated to each vertex identifies the timelike transverse direction to each tetrahedron.  Forming the appropriately parallel transported scalar products among these directions we can build the matrix of cosines of the hyperdihedral boost angles $\Theta_{ab}$.  This Gram matrix fully encodes the geometry of a unique 4-simplex. One finally checks the consistency of the reconstructed geometry with the other data contained in the holonomies, such as the areas $\mathrm{a}_{ab}$ of the triangular faces. The determinant of the Gram matrix encodes the sign of the curvature. 

\emph{Chern-Simons quantization}---Having translated the space of four-dimensional simplicial geometries into the space of flat connections on a three-manifold, it becomes natural to quantize this space using Chern-Simons (CS) theory. Indeed, CS theory is the only three-dimensional gauge theory with flat connections as solutions to the equations of motion. To connect with the geometrical reconstruction above we still miss one ingredient, the geometricity constraint. We impose this constraint as a boundary condition on the CS path integral on $M_3$. 

 Both for mathematical convenience and with the intention of eventually coupling this theory to fermions, we lift the Lorentz group to $\mathrm{SL}(2, \mathbb{C})$. With $I[A]  = \frac{1}{4 \pi} \int_{M_3} \mathrm{tr} \left[ A \mathrm{d} A +\frac{2}{3} A^3\right]$ the complex action is, \cite{Witten:1991, BarNatan:1991},
 \begin{equation}
 S[A,\bar{A}] = \frac{k}{2} I[A] +\frac{\bar{k}}{2} I[\bar{A}],
 \end{equation}
 where $A$ is an $\mathrm{SL}(2, \mathbb{C})$ connection in the fundamental representation, $k$ is  the complex level (or inverse coupling), and the overbars indicate complex conjugation. 
 Efforts to define precisely the $\mathrm{SL}(2,\mathbb{C})$ theory have, in part, relied on analytic continuation of $\mathrm{SU}(2)$ CS theory \cite{Witten:2010}, see also \cite{Dimofte:2015,Andersen:2014}. For our present purposes, the formal path-integral definition is sufficient.  

With quantization in mind we seek a set of canonically conjugate coordinates in the classical theory. A standard and convenient choice are complex Fenchel-Nielsen (FN) coordinates $x$ and $y$, which can be understood as follows \cite{Kabaya:2011,Veen:2013,HHKR:2015}. We restrict attention to the holomorphic sector, with the other one obtained by complex conjugation. The Atiyah-Bott symplectic structure, $\Omega$, of CS theory sets transverse and longitudinal parts of the connection as conjugate. A symmetry reduction of the gauge trades the connection variables for our transverse and longitudinal holonomies \cite{Goldman:1984}, which also turn out to be conjugate. More precisely, if we introduce the eigenvalues of the transverse and longitudinal holonomies
\begin{equation}
\label{FNdefs}
x_{ab} = e^{u_{ab}} \qquad \text{and} \qquad y_{ab}=e^{-\frac{2 \pi}{k} v_{ab}}
\end{equation}
respectively, then  $\{ u_{ab} , v_{cd} \} = \delta_{ab,cd}.$ To complete the variables introduce a trinion (pants) decomposition of each 4-valent vertex and the associated $\{ u_a, v_a \}_a$. 

The variables $\{ u_a, v_a \}_a$ are purely imaginary as the geometricity constraint imposes that they are in an $\mathrm{SU}(2)$ subgroup of $\mathrm{SL}(2, \mathbb{C})$. The reconstruction theorem implies that $u_{ab} \propto i \Lambda \mathrm{a}_{ab}$, where $\mathrm{a}_{ab}$ is the area of the corresponding triangle in the 4-simplex, and hence $u_{ab}$ is also purely imaginary. Finally, $\frac{2 \pi}{k} v_{ab}$ is complex and its real part is $\propto \Theta_{ab}$, which is the hyperdihedral boost angle between tetrahedra $a$ and $b$. All of the logarithmic coordinates $u$ and $v$ have the standard branch ambiguity, leaving their values defined only up to an element of $2 \pi i \mathbb{Z}$ and this will play a role briefly. 

With the classical coordinates specified we turn to constructing the physical wave function of a quantum grain of geometry. We do this in the $u$-representation and for notational convenience  introduce $u = (u_a, u_{ab})$.  The wave function is constructed using the path integral 
\begin{equation}
Z(u, \bar{u} ) = \int_{u, \, \bar{u} } \mathcal{D} A \mathcal{D} \bar{A} \; e^{i S[A, \bar{A}]}\, ,
\end{equation}
where the subscripts indicate that the integration is performed at fixed $u$. This wavefunction satisfies a set of Schr\"odinger-type equations, $\mathbf{A}(\hat{x}, \hat{y};k) Z(u, \bar{u}) = 0 $ and its anti-holomorphic analog,
which are quantum mechanical implementations of the topological constraints of Eq. \eqref{eq_topoconstr}. As functions of the FN coordinates, the $\mathbf{A}$ are known as $A$-polynomials, and are a specific quantization of the celebrated topological invariant of 3-manifolds \cite{Cooper:1994,Gukov:2005, Dimofte:2013}. The wave function $Z(u,\bar{u})$ factorizes, \cite{Beem:2014},  
\begin{equation}
\label{wavefunc}
Z(u,\bar{u}) =  \text{\Large $\Sigma$}_{\alpha} n_{\alpha} Z^\alpha(u) Z^{\alpha}(\bar{u}).
\end{equation}
Here, $Z^\alpha(u)$ and $Z^{{\alpha}}(\bar{u})$ are the holomorphic and anti-holomorphic 3d-blocks, and $\alpha$ labels different branches of the solutions to the $A$-polynomial equations. The reconstruction demonstrates that there are two branches, $ \alpha =1, 2$, corresponding to the two parities of the 4-simplex. In terms of FN coordinates these solutions differ only in the sign of the real part of $\frac{2\pi}{k} v_{ab}$, that is, in the sign of the boost $\Theta_{ab}$. Requiring suppression of other possible branches can be used to fix the coefficients $n_{\alpha}$ and is under study. 

Note that the holomorphic block $Z^\alpha(u)$ is defined through an analytic continuation of the $\mathrm{SU}(2)$ CS theory, hence the name `holomorphic'  \cite{DimofteGukov:2009,Witten:2010,Dimofte:2013,Beem:2014}. While the Darboux coordinates $u$ and $v$ are useful for the WKB analysis below, it turns out that non-perturbatively the holomorphic block is a meromorphic function of $x$ and thus must be periodic in $u$. This is connected with the finiteness of our model discussed in the conclusion.

\emph{WKB analysis}---To understand the semiclassical, small $\hbar$, limit of these wavefunctions we use a geometrical approach to the WKB approximation. A WKB wavefunction is of the form $\text{\Large $\Sigma$}_{\alpha} R_{ \alpha } e^{\frac{i}{\hbar} I_{ \alpha }}$. The phase function $I_{ \alpha }$ is a solution of the Hamilton-Jacobi equation and is most easily constructed using the geometry of Lagrangian submanifolds of the phase space \cite{Littlejohn:1992}. Here we focus on the relative phase between the two branches of Eq. \eqref{wavefunc}. We comment on the overall phase below. 

\begin{figure}[t] 
   \centering
   \includegraphics[width=2in]{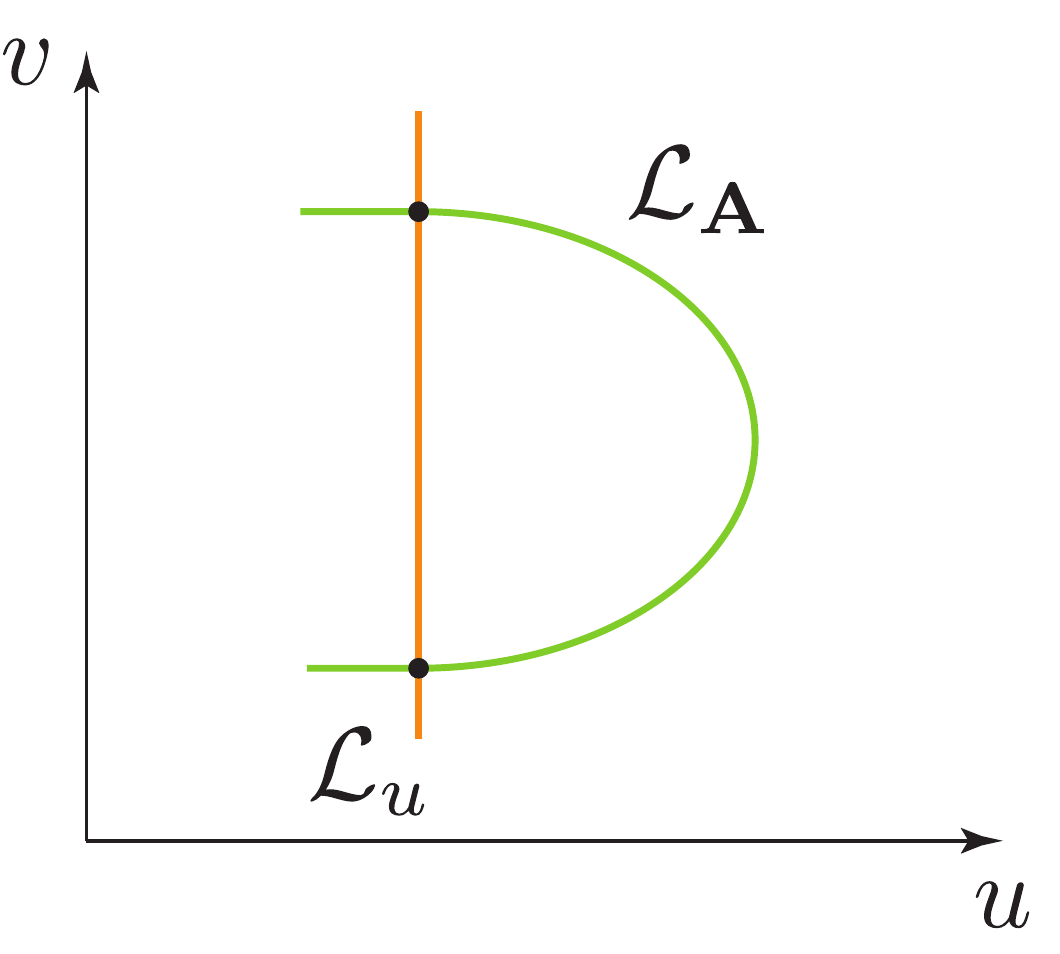} 
   \caption{A sketch of the Lagrangian manifolds $\mathcal{L}_{\mathbf A}$ and $\mathcal{L}_{u}$ and their intersections in $uv$-space.}
   \label{Lagrangian}
\end{figure}
To calculate the phase difference $I_{12} \equiv I_2-I_1$, we have to choose a loop $\ell$ in phase space connecting the two critical points and integrate the Liouville form along it, $I_{12}=\int_{\ell} \vartheta$. For the result to be a solution of the Hamilton-Jacobi equation, the loop is chosen to lie along two Lagrangian submanifolds. The first one, $\mathcal{L}_{\mathbf{A}}$, is implicitly defined by the equations of motion $\mathbf{A}=0$. The second, $\mathcal{L}_{u}$, fixes the representation of the wavefunction and is simply the plane $u = \text{const.}$ These two manifolds intersect precisely at the stationary phase points, where a pair of parity related classical geometries is picked out, see Fig. \ref{Lagrangian}. 

For calculating the integral of the Liouville form the choice of path on a Lagrangian manifold is locally irrelevant due to the defining property $\Omega|_{\mathcal{L}}=0$.   One remaining subtlety is the shift of the phase as you integrate through a caustic of $\mathcal{L}_{\mathbf{A}}$, where the geometry is degenerate, this is the well-known Maslov index $\eta$ \cite{Berry:1972,Esterlis:2014}. We defer study of this index, which may vanish \cite{Witten:1991,BarNatan:1991,Gukov:2005}, to future work. 

An effective technique to calculate $I_{12}(u)$ is to consider small deformations of the loop. To this end, we consider two nearby solutions differing by a small amount $\delta u$. The variation $ \delta I_{12}$ splits into contributions along $\mathcal{L}_{u}$, $\mathcal{L}_{u+\delta u}$, and $\mathcal{L}_{\mathbf A}$. The first two contributions vanish because $\vartheta = v\mathrm{d} u$. Thus, 
\begin{equation}
\delta I_{12} = (v^{(2)}  - v^{(1)}) \delta u = \frac{\Lambda k}{12 \pi i} \text{\Large $\Sigma$} \Theta_{ab}  \delta \mathrm{a}_{ab}
\end{equation}
and is calculated along the manifold of geometrical solutions $\mathcal{L}_{\mathbf{A}}$. The $ \delta u_{a}$ contributions cancel between the two branches. Hence this has exactly the form of the Schl\"afli variation \cite{Schlafli:1858, Haggard:2015}. The Schl\"afli identity then allows us to conclude that
\begin{equation}
I_{12} = \frac{ \Lambda k}{12 \pi i} \left( \text{\Large $\Sigma$} \Theta_{ab} \mathrm{a}_{ab} - \Lambda V + C_{12}\right) ,
\end{equation}
where $C_{12}$ is a geometry independent  integration constant, possibly including $\eta$,  that we drop in what follows.  The geometric origin of the Schl\"afli identity guarantees that the sign of the $\Lambda$ in the action agrees with that of the reconstructed 4-simplex. Restricting to the real contour of functional integration, where $A$ and $\bar A$ are complex conjugated variables, we obtain
\begin{equation}
\label{ZCS}
Z \sim \cos\left\{  \frac{  \Lambda \Im (k)}{12 \pi \hbar}   \left[\text{\Large $\Sigma$} \Theta_{ab} \mathrm{a}_{ab} - \Lambda V \right] \right\},
\end{equation}
up to a global phase, which simplifies in the case of 4-simplices in the bulk of the triangulation.
Identifying the coefficient $  \frac{ \Lambda \Im(k)}{12 \pi \hbar} $ with $(8 \pi G \hbar)^{-1}$ we recognize the curved Regge action calculated on our semiclassical grain of spacetime. This result is analogous to the asymptotics of the Turaev-Viro model for three-dimensional quantum gravity.

While discussing the branches of $\mathcal{L}_{\mathbf{A}}$ we ignored the lift ambiguity of the logarithmic FN coordinates. When properly accounted for, this ambiguity enters the argument of Eq. \eqref{ZCS} and adds a term  
\begin{equation}
\label{Nint}
 \frac{ \Lambda \Re(k)}{6 \hbar}  \text{\Large $\Sigma$} N_{ab} \mathrm{a}_{ab},
 \end{equation}
 where the $N_{ab}$ are arbitrary integers. Therefore to have an unambiguous state we require this expression to be an integer multiple of $2 \pi$. This forces a quantization of the areas in units of $8 \pi \gamma G \hbar$, where $\gamma = \Im(k)/\Re(k)$. 
 
 This concludes the WKB analysis of a single 4-simplex. Just as a simplicial complex can be built from gluing individual 4-simplices we can glue graph complement manifolds using surgery at the graph vertices; the geometricity conditions must be imposed at all surgery sites. At fixed areas the WKB analysis goes through simplex by simplex and gives a quantum simplicial complex weighted by its global curved Regge action. A full understanding of the physics on the whole complex and its continuum limit is involved. For example,  what happens when you sum over the areas (the flatness problem) and the role of the local fluctuation of a simplex's parity (the cosine problem) are recognized difficulties under investigation. 
 
 Crucially, although the geometrical reconstruction proceeds simplex-by-simplex, the cosmological constant and its sign must be semiclassically consistent across the complex. A WKB solution exists only if the curvature of all the simplices is the same. Note that this does not mean that the simplicial manifold cannot support curvature defects as in Regge calculus. This consistency is a consequence of the fact that the sign of the curvature is already determined at the level of the tetrahedra in the boundary of the simplex and therefore propagates to neighboring simplices. 
 
\emph{Relation with the EPRL model---}A natural way to impose boundary conditions on the connections in $M_3=S^3 \setminus \Gamma_5$ is to insert an appropriate $\Gamma_5$ Wilson graph operator in $S^3$. A Wilson graph operator satisfying our desiderata can be built out of the EPRL intertwiners at the core of spinfoam quantum gravity \cite{EPR:2007,EPRL:2008}. These desiderata are the geometricity constraints and their consistent implementation with quantized areas. In the spinfoam context the geometricity constraints are referred to as `simplicity constraints' and are used to turn topological $BF$ theory into general relativity by imposing that the $B$-field is the square of the tetrad field, $B = e \wedge e$.  

The area spectrum follows from the quantization of the classical variables via $\mathrm{SU}(2)$ and $\mathrm{SL}(2, \mathbb{C})$ representation theory. This procedure associates the spin quantum numbers of $\mathrm{SU}(2)$ to geometrical areas. The spins of $\mathrm{SU}(2)$ are half-integers, not integers, and this introduces an area-dependent sign in Eq. \eqref{ZCS}, already present in the $\Lambda=0$ spinfoam asymptotics \cite{Barrett:2010} and in the Ponzano-Regge model \cite{Barrett:2009}. The spectral spacing is determined by the Planck area, $8\pi G \hbar$, multiplied by the dimensionless Barbero-Immirzi parameter,  and the latter can be identified with our $\gamma$. 

Early work \cite{Ashtekar:1986} on loop gravity attempted to quantize the theory with $\gamma=i$ but ran into difficulties \cite{Barbero:1995}. Analytically continued CS theory allows for a continuation of both $\Re(k)$ and $\Im(k)$ to arbitrary complex numbers \cite{Witten:2010}. This freedom provides an opportunity for spinfoam models, since it might be used to consistently continue the real Barbero-Immirzi parameter back to the imaginary unit. 
This continuation is related to the original self-dual Ashtekar variables, with multiple advantages such as a more intimate relation to covariant geometries, a simplification of the dynamics, and compelling black-hole entropy counting \cite{Frodden:2014,Carlip:2015}. 

The present work gives a CS theory generalization to four dimensions of the quantum group deformation used in three-dimensional quantum gravity models to implement a non-vanishing cosmological constant.  Although for three-dimensional models finiteness on a fixed simplicial complex is limited to the positive cosmological constant case \cite{Turaev:1992}, in the model presented here finiteness extends to both signs of $\Lambda$. In fact, finiteness is a consequence of the presence of a finite number of states, see \cite{Han:2011, Fairbairn:2012}, which in turn follows from the periodicity and discreteness of $u$, Eqs. \eqref{FNdefs} and \eqref{Nint}. This illustrates one of the, we hope, many advantages of including a cosmological constant in quantum gravity.\\

HMH acknowledges support from Bard College and warm hospitality at the CPT, Aix-Marseille University.  MH acknowledges support from the Alexander von Humboldt Foundation. This research was supported in part by Perimeter Institute for Theoretical Physics. Research at Perimeter Institute is supported by the Government of Canada through Industry Canada and by the Province of Ontario through the Ministry of Research and Innovation.


\providecommand{\href}[2]{#2}\begingroup\raggedright

\end{document}